**Fractal-like plasmonic self-similar material with a tailorable plasma frequency in the near-infrared**

*Denis Garoli[1], Eugenio Calandrini[1], Angelo Bozzola[1], Andrea Toma[1], Sandro Cattarin[2], Michele Ortolani[3] and Francesco De Angelis[1]\**

[1]Istituto Italiano di Tecnologia, via Morego 30, I-16163, Genova, Italy.
[2] Istituto di Chimica della Materia Condensata e di Tecnologie per l'Energia (CNR-ICMATE), Corso Stati Uniti 4, I-35127 Padova, Italy.
[3]Dipartimento di Fisica, Sapienza Università di Roma, Piazzale Aldo Moro, 5, I-00185 Roma, Italy.

E-mail: Francesco.deangelis@iit.it

**Abstract**

In this work, we show that modulating the fractal dimension of nanoporous gold allows its effective dielectric response to be tailored over a wide spectral range of infrared wavelengths. In particular, the plasma edge and effective plasma frequency depend linearly on the fractal dimension, which can be controlled by varying the pore and ligament sizes. Importantly, the fractal porous metal exhibits superior plasmonic properties compared to its bulk counterpart. These properties, combined with a longer skin depth on the order of 100-200 nm, enables the penetration of optical energy deep into the nanopores where molecules can be loaded, thus achieving more effective light-matter coupling. These findings may open new pathways to engineering the optical response of fractal-like or self-similar metamaterials without the need for sophisticated lithographic patterning.

1. Introduction

During the last decade, plasmonics has been proposed for applications in several fields from biosensing to solar cells and electrochemistry, and research regarding new plasmonic nanostructures and materials has been continuously growing[1]. The engineering of plasmonic nanostructures enables the control of light at the nanoscale level and the concentration of optical energy into hotspots where the radiation electric field is strongly enhanced when compared to the maximum value attainable in a diffraction-limited focal spot size[2–6]. The hotspots can be engineered down to the size of a few nanometres; thus, these hotspots are comparable to the size of single molecules, which is a detection limit that has previously been demonstrated[7]. Moreover, placing a molecule into a plasmonic hotspot makes it possible to investigate physical and chemical processes featuring high optical energy thresholds, which are otherwise challenging to surpass[8-11]. However, achieving the simultaneous spatial co-localization of nano-objects and enhanced optical energy is extremely challenging due to the intrinsically small modal volume of plasmonic hotspots[11-15]. A typical example is represented by nanosensors working in a liquid solution where the so-called *diffusion limit* often hinders their practical applications to highly diluted samples[16]. The co-localization problem is even more serious in the mid-infrared (mid-IR) region (wavelengths of approximately 2 to 12 microns), where plasmonics can help obtain new enhanced vibrational spectroscopic schemes[17-20] and electromagnetic energy concentrations at frequencies resonant with molecule-specific dipolar vibrations. Moreover, mid-IR plasmonics can be beneficial for future applications because of the recent commercialization of quantum cascade lasers[21] broadly tunable in the mid-IR range, enabling automated lab-on-chip vibrational spectroscopy[22]. Finally, polariton excitations of 2D materials and hybrid nanostructures[23] often fall in the mid-IR region, hence making this range a potential future avenue for novel photonic devices. Accordingly, new valuable materials for IR plasmonics can significantly impact the community, and although progress has been made using non-metallic conductors[19, 24-25], the search for alternative IR plasmonic materials is still an open issue. In this regard, nanoporous gold (NPG) has been widely investigated in the visible / near-infrared (NIR) spectral range[26-27] and recently also in the mid-IR region, where it was shown to boost the optical energy transfer from the far field to molecules adsorbed in the nanopores, thus bypassing the issue of co-localization[28].

In this work, we show that NPG behaves like an optical metamaterial whose effective plasma frequency can be rationally engineered in the mid-IR spectral range by varying the fractal dimension. The plasma edge can be controlled by adjusting the level of porosity. Thus, both the real and imaginary parts of the permittivity can be tailored over a wide range and still maintain a superior quality factor with respect to conventional semiconductor metamaterials or nanostructured metallic surfaces. Importantly, the optical properties in the mid-IR region stem from the intrinsic material features achieved by bottom-up material growth processes; hence, sophisticated lithographic patterning is not required. Under this condition, as verified by numerical simulations, the optical energy can now penetrate deep into the metal structure, where it can be accumulated into the empty spaces of the nanopores. Loading NPG structures with molecules therefore provides a straightforward method for the co-localization of molecules and plasmonic hotspots, which boosts optical energy transfer.

Finally, we suggest that the fractal dependence of the optical properties may be generally valid, and thus can be applied to other nanoporous materials.

2. Results

NPG samples were prepared following the procedure described in the Methods section. Scanning electron microscopy (SEM) images of four exemplificative samples are reported in **Figure 1(a)**. With the naked eye, the void size between the gold ligaments clearly increases with increasing dealloying time from sample A to sample D. The gold filling factor $f$, estimated with a pixel count method, is $f$=0.39 for sample A and slightly decreases to $f$=0.36 for sample D. Therefore, the filling factor is not the main parameter affected by the dealloying procedure. Spatial discrete Fourier transform (FT) of the SEM images (one example is shown in **Figure 1(b)**) provides a compact description of the ligament structure, which cannot be obtained from real-space image inspection. Assuming that the ligament has an ellipsoidal shape, the inverse standard deviation of a two-dimensional Gaussian surface fitted to the FT image provides an estimate of the average ligament size. With this approach, it is now possible to describe the difference between samples A to D: as shown in **Figure 1(c)**, the estimated ligament area $1/\sigma^2$ increases linearly with dealloying time. Combined with an almost constant filling factor, this fact indicates that the density of individual ellipsoidal ligaments is inversely proportional to the dealloying time (**Figure 1(a)**).

The reflectance of NPG films grown on silicon substrates was determined by Fourier transfer infrared (FTIR) spectroscopy at quasi-normal incidence (**Figure 1(d)**). The increasing pore size produces a progressive shift in the plasma edge, namely, the frequency below which a conducting material exhibits totally reflective behaviour. At frequencies above the plasma edge, the skin depth $\delta$ starts to increase with frequency, and the electric field can penetrate deeply inside the material. At some slightly higher frequencies, usually defined as the plasma frequency $\omega_p$, the dielectric permittivity $\varepsilon'(\omega)$ crosses zero, and the material no longer behaves like a conductor. The frequency range where penetration is high but the dielectric permittivity ε' is still negative can be defined as the range of truly plasmonic behaviour. For bulk gold, the range above the plasma edge (at approximately 15000 cm⁻¹) corresponds to a frequency range of strong absorption due to interband transitions (high imaginary part of the permittivity ε"); therefore, bulk gold cannot be used for efficient plasmonic energy transfer from radiation to molecules. Instead, our NPG samples feature plasma edges between 4000 and 2000 cm⁻¹, which are far from the interband transitions of gold.

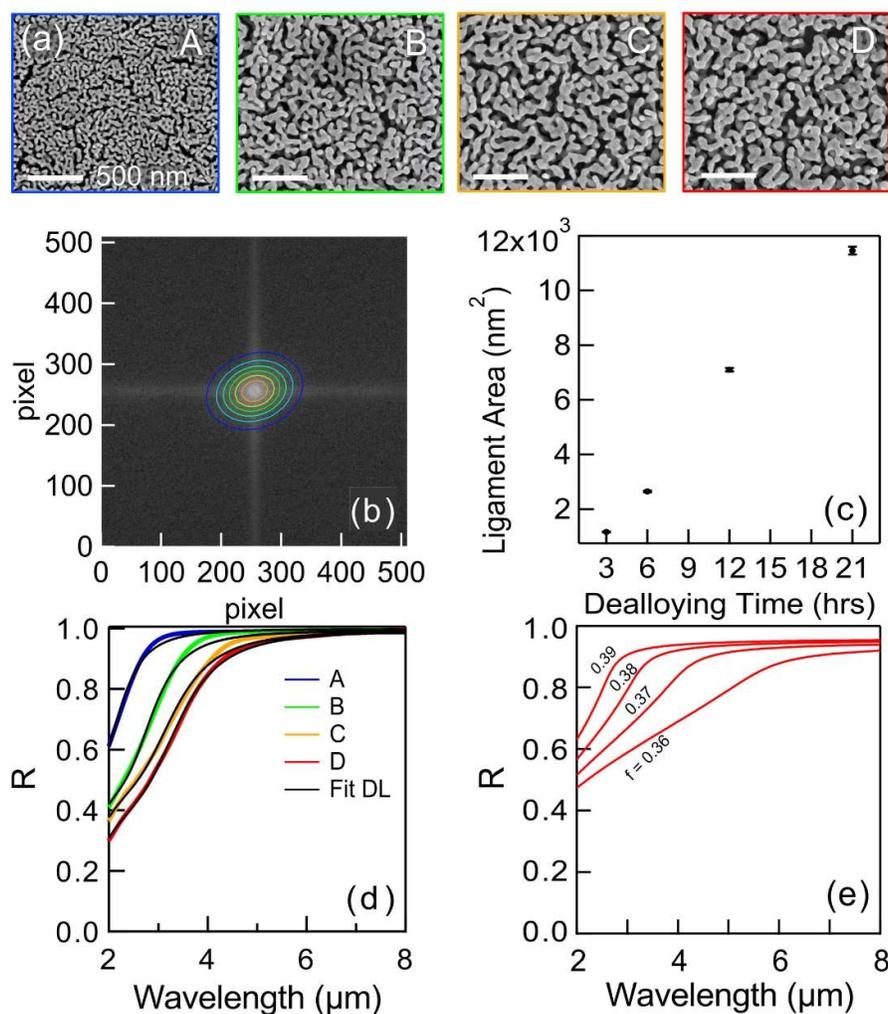

**Figure 1.** *Tuning the permittivity of porous gold with the dealloying time. (a) SEM images of the investigated NPG samples obtained with different dealloying times: 3 hours (blue - A), 6 hours (green - B), 9 hours (orange - C), and 12 hours (red - D). (b) FT of the SEM images. (c) Ligament area vs. dealloying time. (d) Measured FTIR reflectance spectra for samples of thick nanoporous metal films obtained with different dealloying times. The corresponding effective plasma frequencies obtained with the Drude-Lorentz model are 6200, 5000, 4040 and 3840 cm-1 for samples A, B, C and D, respectively. (e) Reflectance calculated with the*

*Bruggeman Model for the effective-medium dielectric constant using the filling factors obtained with the pixel count method and indicated in the figure.*

The optical properties of inhomogeneous media such as NPG can be modelled in principle within the effective-medium approximation (EMA), in which the two known dielectric functions of two starting materials (in this case, vacuum and bulk gold) are combined in an analytical relation that defines the effective dielectric function. According to the different hypotheses that can be formulated on the inhomogeneous material structure, different assumptions can be made regarding the local depolarization fields. The Maxwell-Garnett EMA[29] typically works well for isolated particles of one strongly absorbing material dispersed in a weakly absorbing continuous matrix; this is clearly not the case for NPG. In the Landau-Lifshitz-Looyenga approach[30], the connectivity among the ligaments becomes the key parameter; however, this works at the filling-factor limit, which is also not the case for our samples. The Bruggeman formulation of the EMA[29], where the mutual depolarization fields are considered in a mean-field approach without any assumptions regarding the material properties, better fits our needs of modelling NPG. However, the vacuum does not support the polarization charge densities but only the polarization fields. The formulation of the Bruggeman EMA used here is:

$$f_m \frac{\varepsilon_m - \varepsilon_{eff}}{\varepsilon_m + 2\varepsilon_{eff}} + f_0 \frac{\varepsilon_0 - \varepsilon_{eff}}{\varepsilon_0 + 2\varepsilon_{eff}} = 0 \tag{1}$$

where $\varepsilon_m$ is the dielectric function of gold, $\varepsilon_0$ is the vacuum dielectric constant, $\varepsilon_{eff}$ is the resulting effective dielectric function of the nanoporous material, and $f$ is the filling factor, which is the only parameter through which the sample structure enters the formula. In **Figure 1(e)**, the reflectivity calculated from $\varepsilon_{eff}$ is shown for all samples. Clearly, while the other two considered methods completely fail (see supporting information), the Bruggeman EMA captures the main physical phenomenon at work in NPG (depolarization fields and charges), as it can qualitatively explain the redshift of the plasma edge but cannot quantitatively describe the electrodynamic response of the material because the complex material structure is not described by the filling factor alone.

Due to the general difficulty in developing a coherent EMA for NPG, we analysed the electrodynamic response of NPG by using the phenomenological Drude-Lorentz model for the complex permittivity $\varepsilon(\omega)$. In NPG, the effective density $n$ of free carriers, defined as those electrons that contribute to currents down to the zero-frequency limit, as described by the Drude model, is reduced compared to that in bulk gold because of the presence of surface boundaries and finite-size structures providing shape resonances in the infrared (IR) range. The response of finite-size structures can be effectively represented by a finite-frequency IR conductivity peak centred in the mid-IR region (Lorentz term representing localized carriers). These structures result in high-frequency depolarization phenomena that screen the free-carrier response. This response results in a redshift in the plasma edge and a reduction in the effective $n$. Note that the finite-frequency IR conductivity peak is different from the high-frequency depolarization usually accounted for by introducing an infinite dielectric constant $\varepsilon_\infty$. Instead, this peak represents not only interband optical transitions at photon energies in the visible/UV range, outside the IR measurement range, but also those present in bulk gold. The resulting expression is:

$$\varepsilon(\omega) = \varepsilon_\infty - \frac{A_D}{\omega^2 + i\omega\gamma_D} + \frac{A_L}{\pi} \cdot \frac{\gamma_L/2}{(\omega - \omega_L)^2 + (\gamma_L)^2} \qquad (2)$$

where $A_L$, $\omega_L$, and $\gamma_L$ are the strength, frequency and width of the Lorentz term, respectively, and $A_D$, and $\gamma_D$ are the intensity and width of the Drude term, respectively. The expression of **Eq. 2** is fitted to the reflectivity data of each sample with a Kramers-Kronig consistent routine[31], and the fitting results are shown in **Figure 1(d)**. The values of the fitting parameters are reported in the Supplementary Information file. In **Figure 2(a)-(c)**, the corresponding real and imaginary parts of the dielectric function and the skin depth are plotted. The most relevant parameter in **Eq. (2)** is the Drude weight $A_D$, which is related to the physical parameters by the classical Drude relation $A_D = \sqrt{ne^2/m}$, where $n$, $e$, and $m$ are the effective free carrier density, charge, and mass (considered constant in the different samples), respectively. The key quantity for IR plasmonic applications, however, is not the Drude weight but rather the effective plasma frequency, i.e., the frequency where $\varepsilon'(\omega_p)=0$. In the presence of Lorentz oscillators, there is no simple analytic relation between the Drude weight and the plasma frequency, although they are normally interrelated by a particular monotonic function.

In NPG, the free carrier charge and mass, which are truly microscopic quantities, are considered equal to those of bulk gold. On the other hand, the density $n$ determined from $A_D$ will be different in each NPG sample because each sample morphology corresponds to a Lorentz conductivity peak at a different IR frequency with a different strength and width. The effective plasma frequency $\omega_p$ depends on all these parameters but is found to be related to the actual film morphology in a direct way, as explained below. Both the Drude weight and the effective plasma frequency decrease significantly for NPG samples prepared with longer dealloying times (see Table 1). The shift in the effective plasma frequency is therefore related to the evolution of the morphology properties of the NPG film under the dealloying process and not to the gold filling factor *f*, which is approximately constant. Physically, the redshift in the plasma edge and the decrease in $\omega_p$ with increasing ligament size with a constant filling factor could be interpreted as a reduction in the effective free carrier density in a volume with dimensions on the order of the wavelength, which is similar to what was suggested years ago at the beginning of the field of metamaterials and in studies related to spoof plasmons[32-33]. However, there is no direct correlation between *n* and the dealloying time, making it difficult to predict the material properties before material growth.

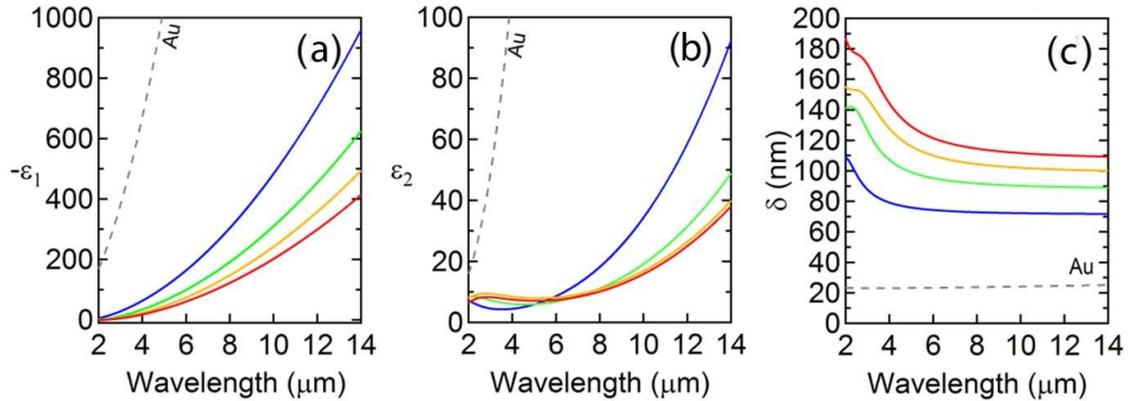

**Figure 2.** *Real (a) and imaginary (b) parts of the permittivity and the skin depth (c) estimated with the Drude model for the investigated samples. The reference spectra of bulk gold are also shown with black dashed lines.*

To better understand the link between the morphology and electrodynamic properties and to provide a deterministic formulation for plasmonic material engineering, we investigated the morphology of the porous structures using fractal theory. This theory is a very powerful method for relating nanoscale morphology and macroscopic responses[34]. Again, we used real-space SEM image analysis for extracting quantitative information regarding the 3D structure starting from their 2D greyscale representation[35]. The fractal analysis retrieves the so-called fractal dimension of the system under investigation. This quantity is the non-integer counterpart of the dimension concept in Euclidean geometry and characterizes the percolation properties of the porous structure. Intuitively, if the fractal dimension approaches 3, then the film properties approach those of bulk gold. The fractal dimension is estimated using the box-counting method, which assigns a 0 or 1 value to each pixel in the SEM image. The discretized SEM images can be divided into self-similar smaller squares, each with its characteristic length *L*. The density $N(L)$ of squares of size $L$ is computed explicitly. The fractal dimension $D_f$ can then be calculated according to the equation:

$$D_f = \frac{\ln N(L)}{\ln L}.$$ (3)

The power-law relation between N(L) and L, which defines $D_f$, is shown in Figure 3(d) for one of the samples. In Figure 3(e), the effective plasma frequency $\omega_p$ and the effective free carrier density $n$ are shown as a function of $D_f$. The points for bulk gold (integer dimension equal to 3), corresponding to its plasma frequency are also shown in the same plot. The linear correlation is significant, given the small number of data points. We deduce that the fractal dimension, which can be directly computed from the SEM images, is the key morphological parameter to predict the plasmonic properties of NPG and simply tune them at will with the dealloying time. Notably, considering the lower values of the fractal dimension, it seems to be possible to drive the plasmonic properties down to the terahertz spectral range.

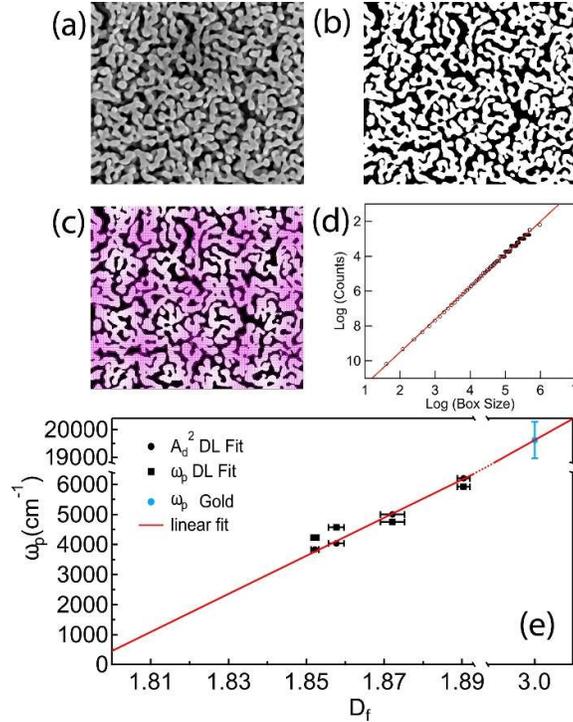

**Figure 3. Fractal analysis.** *(a) Original image. (b) Binarized image. (c) Example of one step of the box-counting method. (d) Application of the box-counting method. (e) Connection between the effective Drude model and the fractal analysis.*

Because of the increased dealloying time, in addition to the redshift of the plasma edge and $\omega_p$, the values of the real (Figure 2(b)) and imaginary (Figure 2(c)) parts of the effective dielectric function of NPG in the IR range became much lower than those of the bulk[36] (black curves in Figure 2(b)-(d)). The lower $\varepsilon_1$ and $\varepsilon_2$ values imply a reduction in the extinction coefficient $\kappa$ =$\left\{\frac{1}{2}(\varepsilon_1^2 + \varepsilon_2^2)^{1/2} - \frac{\varepsilon_1}{2}\right\}^{1/2}$ and therefore an increase in the skin depth $\delta=c/\omega\kappa$ when compared to bulk gold. This is a second key feature of NPG, which allows the co-localization of molecules and plasmonic hotspots in the mid-IR region, as we now show.

## 3. Discussion

The observed reduction in the dielectric function with decreasing fractal dimension has important consequences. The lower imaginary part of the dielectric function results in lower ohmic losses per unit volume. The increased skin depth enables the IR radiation to penetrate deep into the buried nanopores. In **Figure 2(c)**, the skin depth is plotted for the different samples and can be tuned in the range of 100-200 nm for comparison with the skin depth of bulk gold, which is on the order of 30 nm[36]. Importantly, the NPG volume in which the field penetrates is not entirely filled with metallic atoms but contain voids up to a volume fraction $1 - f$=0.64. The nanopores buried within a skin depth $\delta$ from the surface can then host both the plasmonic hotspots and molecules of interest. The co-delivery of optical energy and molecules into the same location then boosts the energy transfer from the electromagnetic waves to the molecular vibrations in the mid-IR[28].

To provide a quantitative evaluation of the energy transfer phenomenon, we numerically evaluated the interaction of a dielectric nanoparticle with porous and bulk gold illuminated with IR light. The dielectric nanoparticle represents molecules of interest or other nano-objects. A full 3D electromagnetic calculation that considers the real geometry of the nanopores is currently a difficult challenge that requires a considerable computational effort. Hence, we set up an electromagnetic simulation with Comsol Multiphysics software®, and we treated the porous metal as a uniform material whose permittivity is given by the values experimentally determined by means of FTIR and subsequent Drude-Lorentz modelling. We considered 4 representative cases: a dielectric nanoparticle (diameter of 10 nm, *n=1.5*, *k=1*, *ε=1.25+3i*) buried 50 nm beneath the surface (**Figure 4 (a)**) or placed onto the porous gold surface (**Figure 4 (b)**). In **Figure 4 (c)** and **(d)**, bulk gold is compared to porous gold. The top panels of **Figure 4** show the electric field distribution (cross section) for the four cases considered when the system is excited with a plane wave impinging normally to the surface at λ=2.45 µm. We noticed that the energy transfer is strongest when the nanoparticle is embedded in the porous material.

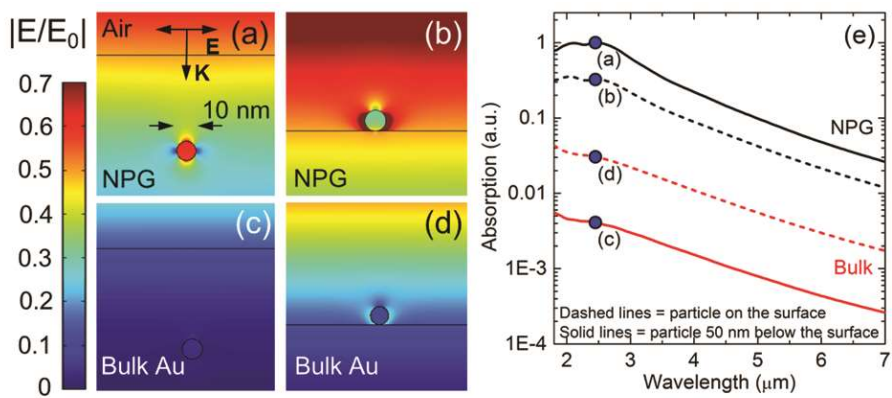

**Figure 4. Electric field penetration and light-matter interaction: NPG vs bulk gold.** *(a-d) Normalized amplitude of the electric field at λ=2.45 µm for a dielectric sphere (diameter=10 nm, n=1.5, k=1, ε=1.25+3i) embedded in the metal (depth=50 nm) or on the surface. (e) Comparison of the absorption spectra of the dielectric sphere for the four configurations.*

Interestingly, the porous material exhibits a more effective energy transfer compared to bulk gold even when the nanoparticle is placed onto the surface (not embedded). This effective energy transfer is due to a strong mid-IR field confinement for NPG in the direction normal to the surface (see supporting information for more details), which is a typical feature of bulk gold in the visible range but is completely lost at mid-IR frequencies. The optimal energy transfer is achieved when the excitation frequency is close to the effective plasma frequency of the porous material, but the advantage of the porous structure over bulk gold extends to the whole IR range considered. The latter fact is detailed in **Figure 4(e)**, where we report the amount of optical energy absorbed by the nanoparticle in the range 2-7 µm. The optical energy delivered to the nanoparticle is at least 10 times and up to 100-300 times higher when NPG is used. Remarkably, this behaviour is widely tunable in the IR range. Furthermore, the very high surface/volume ratio characteristic of porous gold allows the adsorption of large amounts of analytes inside the porous matrix, which unlocks new prospects for biosensing applications, thermoplasmonics, catalysis and enhanced nonlinear optical properties.

4. Conclusion

In summary, we have investigated the mid-infrared optical properties of nanoporous gold films with different void and ligament volumes but almost constant filling factors. We have shown that the effective dielectric function could be described only qualitatively within simplest effective medium approximation theories; hence, we exploited the phenomenological Drude-Lorentz model. We found that key parameters such as the effective plasma frequency and the effective free carrier density are related to the nanoscale material morphology and can be described by fractal theory. Importantly, the effective plasma frequency depends linearly on the fractal dimension of the material, hence allowing the predictable tailoring of the optical constants in the infrared range. These findings may open new pathways for engineering the optical response of metamaterials through self-similar and fractal concepts without the need for sophisticated lithographic patterning. Additionally, both the real and imaginary parts of the permittivity exhibit lower values with respect to those of bulk gold while showing a higher quality factor, namely, better plasmonic properties. Moreover, higher values of skin depth are observed, enabling the accumulation of optical energy into the buried nanopores where target molecules can be loaded. All these properties combine to allow the realization of efficient optical energy transfer to molecular vibrations by the co-localization of molecules and plasmonic hotspots buried below the film surface.

5. Experimental Section

The preparation of NPG structures is based on procedures[26,27] where Ag is selectively leached from gold-silver alloys. Adjusting the fabrication procedure makes it possible to tune the film morphology and the pore size, which in turn deeply affects the optical properties. Such tunability has been extensively investigated in the visible/near-IR range, whereas a full characterization in the mid-IR has not been reported[37]. By further refining the fabrication protocols, we obtained high-quality porous films that show a progressive increase in pore and ligament sizes. The film morphology can be tailored by acting on i) the Ag/Au ratio of the alloy; ii) the nitric acid concentration of the etching bath; and iii) the temperature of the etching bath[27, 38]. Here, thin films of NPG were prepared as follows: rectangular silicon substrates were thoroughly degreased in boiling acetone and dried. A 5 nm Ti layer was deposited on a silicon nitride membrane as an adhesion promoter. Subsequently, a Au layer approximately 100 nm thick was deposited over Ti. Both depositions were performed using a Kenosistek® facility in high vacuum ($1 \times 10^{-6}$ mbar) by means of electron beam (e-beam) evaporation at a deposition rate of 0.1 nm/s. Finally, an approximately 500 nm-thick layer of the $Ag_{75}Au_{25}$ alloy (at%) was deposited in a DC turbo sputter coater (Emitech K575X, Emitech Ltd., Ashford, Kent, UK) using a silver/gold alloy sputtering target, $Ag_{62.3}/Au_{37.7}$ (wt.%), GoodFellow. The sputtering was performed at room temperature under an Ar gas flow at a pressure of $7 \cdot 10^{-3}$ mbar and a DC sputtering current of 25 mA. The composition of the alloy ($Ag_{75}Au_{25}$) was selected on the basis of previous studies[31], which claim that the range of 22–25 at.% Au is optimal. Etching baths were prepared by dilution of a concentrated $HNO_3$ solution (Sigma-Aldrich, ACS reagent 70%). The dealloying process was performed with the same chemical procedure but different durations of the dealloying (see **Table 1**). Each sample was then washed in distilled water (1 hours), dried in a nitrogen stream and stored.

**Table 1.** Summary of the dealloying conditions, obtained plasma frequencies and calculated free carrier densities of the samples.

| Sample id. | $HNO_3$ conc. | Dealloying time (h) | Plasma freq ($\omega_p$) ($cm^{-1}$) | Free carrier (n) ($cm^{-3}$) |
|---|---|---|---|---|
| A | 33% | 3 | 6200 | 5.46E+20 |
| B | 33% | 6 | 5000 | 3.60E+20 |
| C | 33% | 9 | 4040 | 2.87E+20 |
| D | 33% | 12 | 3840 | 2.42E+20 |

Acknowledgements

The research leading to these results has received funding from the European Research Council under the European Union's Seventh Framework Program (FP/2007-2013) / ERC Grant Agreement n. [616213], CoG: Neuro-Plasmonics and under the Horizon 2020 Program, FET-Open: PROSEQO, Grant Agreement n. [687089].

# Supporting Information

**Supporting Note 1: Analyses of the optical parameter trends with respect to the film morphology.**

The obtained optical constant values were used to compute the material dispersion curve. A 1D model comprising a thin layer of nanoporous gold (500 nm) on a Si/Ti/Au substrate was implemented by means of Comsol Multiphysics®. A modal analysis allows the computation of the expected plasmonic modes, and the results obtained with NPG were compared with the same analysis performed on a thin layer of standard gold on a Si/Ti substrate (the optical constants were obtained from supporting ref. 1). In **Supporting Figure 1**, we compare the calculated dispersion curves of the optical modes. For the NPG films, there is a strong reduction in the surface plasmon frequency, which falls in the NIR part of the spectrum, as expected.

We now consider NPG as a plasmonic building block for applications in the NIR and mid-IR. A surface plasmon polariton (SPP) propagating along a single metal/air interface is characterized by two important parameters: the propagation length (defined as the 1/e intensity-decay length along the direction of propagation) and the confinement width (the 1/e field-decay width on each side of the interface).[2] There is a trade-off between these parameters such that a weaker confinement results in a longer propagation length, and vice versa. In general, the trade-off is very sensitive to losses in the metal and decreases sharply with higher damping losses.

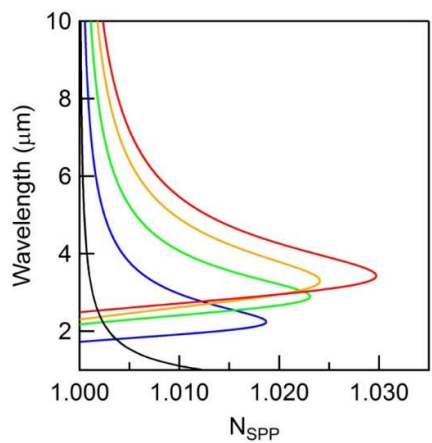

**Supporting Figure 1. Dispersion curves obtained from the experimental dielectric constants.** The black curve corresponds to the case of bulk gold, whereas the coloured curves

denote the different investigated samples of porous gold. The correspondence between colours and preparation conditions is the same as in Figure 2 of the main text.

While the skin depth was already reported in the main text, **Supporting Figure 2** shows the propagation lengths and field confinements for SPPs along the interface between air and NPG. The same analysis is performed for the SPP at the interface between air and bulk gold. Note that NPG provides much better confinement than bulk gold; however, the propagation length for NPG is smaller than that for gold. A series of metrics to determine the efficiency of metals for plasmonics applications are proposed in supporting ref. 2. Although each specific geometry will have a different quality factor, in the limit of low loss and the applicability of electrostatics, two generic limiting cases can be derived: (i) for localized surface plasmon (LSP) applications, and (ii) for extended modes such as SPPs. The obtained dielectric constant results are particularly interesting if we consider the generic expressions for the quality factors for LSP and SPP applications[2]:

$$Q_{LSP} = -\varepsilon_1 / \varepsilon_2, \qquad\qquad Q_{SPP} = \varepsilon_1^2 / \varepsilon_2, \qquad\qquad \text{supporting equation (1)}$$

where $\varepsilon_1$ and $\varepsilon_2$ are the real and imaginary parts of the permittivity, respectively. A comparison between the quality factors obtained with bulk gold and our NPG films is reported in **Supporting Figure 3**. While in most of the cases $Q_{SPP}$ for bulk gold is higher than the respective value for NPG, in the case of sample A, $Q_{SPP}$ is higher for wavelengths above 5.5 μm.

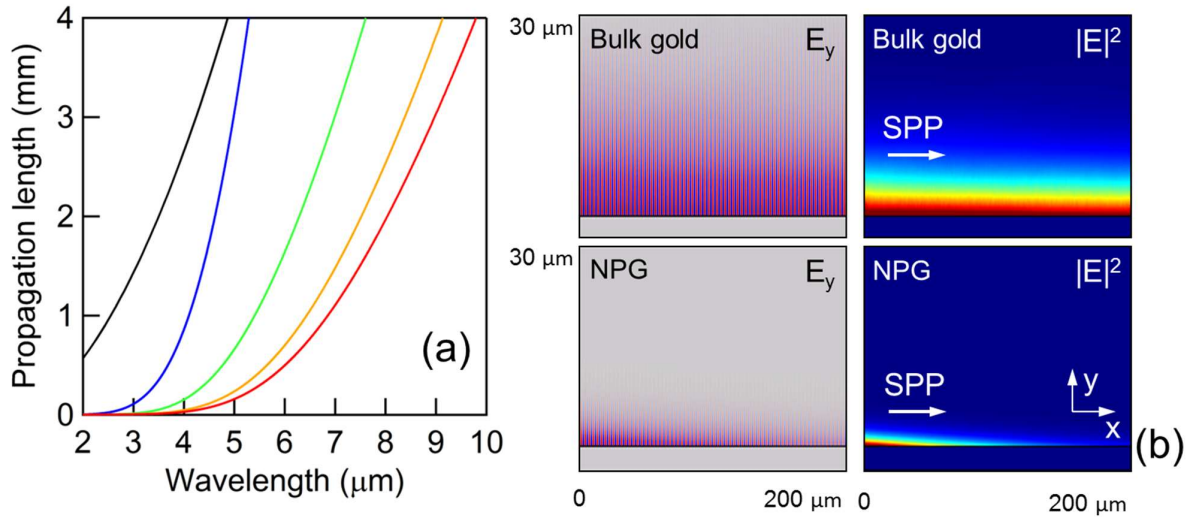

**Supporting Figure 2. (a)** Propagation length for an SPP at the metal-air interface. The black line denotes the case of bulk gold, while the coloured lines represent the different samples of porous gold (the correspondence between colours and preparation conditions is the same as in Figure 2 of the main text). **(b)** Electric field component normal to the metal interface and the squared amplitude of the electric field for an SPP at the metal-air interface. The white arrows denote the injection direction (from left to right). In all the panels, a wavelength of 3 μm is assumed.

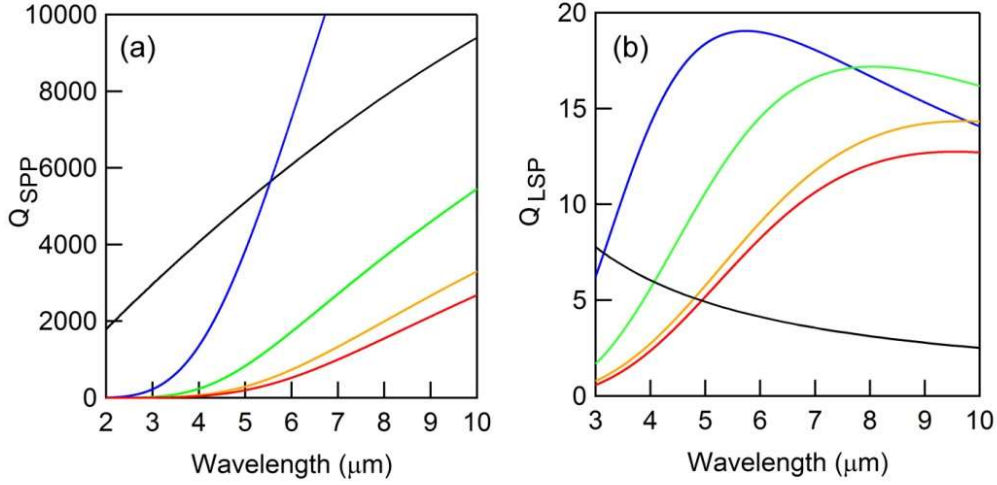

**Supporting Figure 3.** Quality factors for **(a)** generic plasmon propagation and **(b)** LSP excitations.

In contrast, the obtained $Q_{LSP}$ factor is always higher for NPG at wavelengths above 5 μm and reaches approximately 18 at 5.5 μm in sample A. Instead, bulk gold has a $Q_{LSP}$ factor of approximately 5; hence, a 4-fold increase can be obtained between the two materials in the mid-IR region.

**Effective Medium Approximation**

The EMA relies on the process of averaging the properties of the constituents that directly comprise the composite material weighted on their relative fractions. The properties under consideration are usually the conductivity σ or the dielectric function ε of the medium. Here, NPG can be modelled as a mixture of air ($\varepsilon_{h1}(\omega)=1$ and $\varepsilon_{h2}(\omega)=0$) as the dielectric host medium and gold ($\varepsilon_m(\omega)$ as in supporting ref. [3]) as the embedded particles. The relative fraction of the metal in the host is determined by the parameter f. Since the geometrical features of NPG, namely, the pores and the ligaments, are much smaller than the wavelength of the EM wave in vacuum, scattering effects are negligible, and the EMA holds. Several models have been established, and the most prominent ones include the Maxwell-Garnett (MG), the Bruggeman (for isolated particles) and the Landau-Lifshitz-Looyenga (LLL) (for percolating networks) models. In the following sections, these models are described and tested for the system under investigation. The best approximation of the measured reflectance spectra is defined by the relative fraction of metal in air that, within each model, minimizes the following expression:

$$\int \left| R_{ema}(\omega) - R_{\exp}(\omega) \right| d\omega \qquad \text{supporting equation (2)}$$

**Maxwell-Garnett Model**

The MG model considers only the dipolar resonances of small particles highly diluted (or in other words non-interacting) in the host medium[3]. The MG equation reads as:

$$\frac{\varepsilon_{eff} - \varepsilon_h}{\varepsilon_{eff} + 2\varepsilon_h} = f \frac{\varepsilon_m - \varepsilon_h}{\varepsilon_m + 2\varepsilon_h} \qquad \text{supporting equation (3)}$$

where $\varepsilon_{eff}$ is the effective dielectric constant of the medium, $\varepsilon_m$ is the one of the inclusions, $\varepsilon_h$ is the one of the hosts, and f is the volume fraction of the inclusions. The above equation is solved by:

$$\varepsilon_{eff} = \varepsilon_h \frac{2f(\varepsilon_m - \varepsilon_h) + \varepsilon_m + 2\varepsilon_h}{2f(\varepsilon_h - \varepsilon_m) + \varepsilon_m + 2\varepsilon_h} \qquad \text{supporting equation (4)}$$

The dielectric functions provided by this model are shown in Supporting Figure 4 with respect to the f parameter and the calculated normal incidence reflectance. Since the MG model aims at describing a very dilute system, it is inadequate to reproduce the measured spectra. Indeed, the non-percolating hypothesis of this model produces a decrease in the calculated reflectance for longer wavelengths, at variance with the investigated samples.

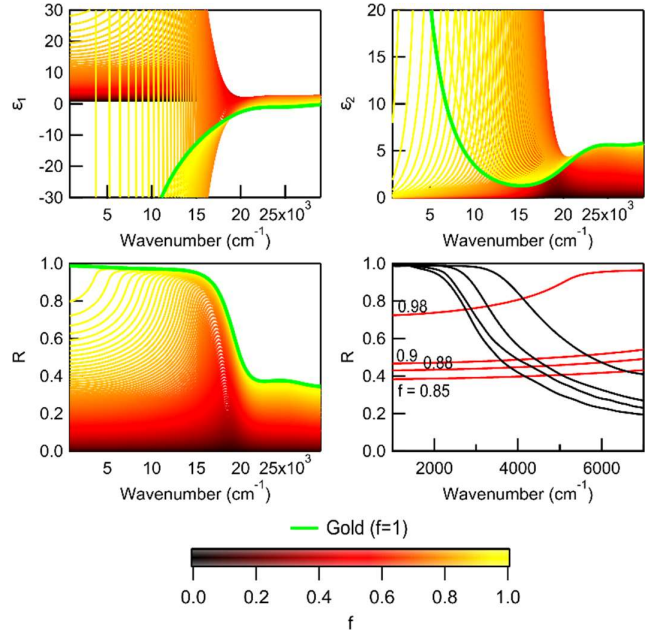

**Supporting Figure 4.** Fit of the optical response by means of the MG EMA model.

**Bruggeman Model**

This model treats the host and the inclusions equally: the system consists of multicomponent particles with different arbitrary dielectric functions[5]. Indeed, the Bruggeman formula is symmetric with respect to interchanging the components, which is attractive for dealing with materials of comparable volume fractions in the mixture. Under this assumption, with the same nomenclature as before, the Bruggeman formula takes the form:

$$f \frac{\varepsilon_m - \varepsilon_{eff}}{\varepsilon_m + 2\varepsilon_{eff}} + (1 - f) \frac{\varepsilon_h - \varepsilon_{eff}}{\varepsilon_h + 2\varepsilon_{eff}} = 0 \qquad \text{supporting equation (5)}$$

In 3D, this equation is solved by:

$$\varepsilon_{eff} = \frac{1}{4}(2\varepsilon_h - 3f\varepsilon_h - \varepsilon_m + 3f\varepsilon_m) \pm \sqrt{8\varepsilon_m \varepsilon_h + (2\varepsilon_h - 3f\varepsilon_h - \varepsilon_m + 3f\varepsilon_m)^2} \quad \text{supporting equation (6)}$$

Supporting Figure 5 shows the dielectric functions provided by this model with respect to the f parameter and the calculated normal incidence reflectance. Although the details of the experimental curves are not reproduced, the plasma edge shape is recovered. Additionally, the equivalence in the treatment of the host and the metallic enclosures of the Bruggeman model demonstrated a more appropriate description in terms of the shape of the calculated reflectance spectra. This fact is also corroborated by the SEM micrographs, where an almost equal composition of pores and ligaments is exhibited.

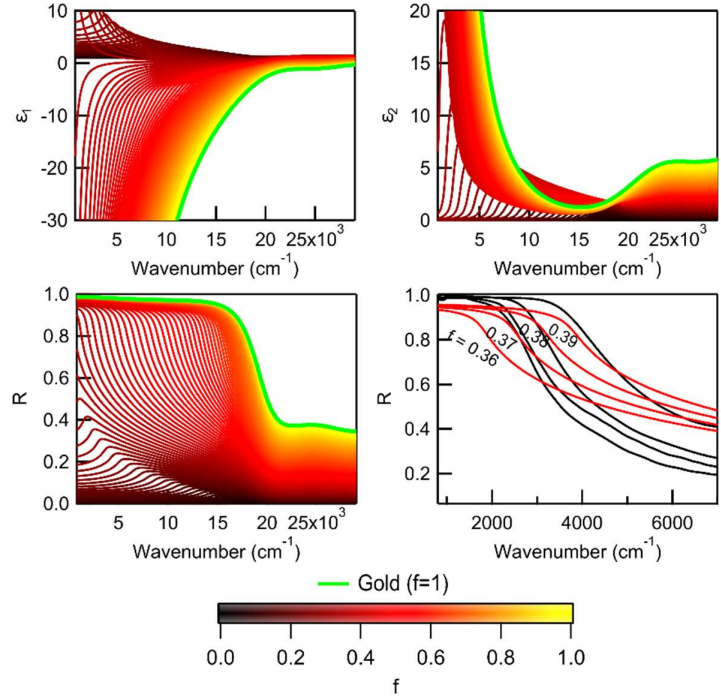

**Supporting Figure 5.** Fit of the optical response by means of the Bruggeman EMA model.

**Landau-Lifshitz-Looyenga Model**

The LLL model performs better than the previous two theories for low volume fraction percolating networks and was independently derived by Looyenga[6] and Landau-Lifshitz. In this model, no shape dependency is considered, and thus, the model is favourably applicable to irregularly shaped particle mixtures. The expression for the effective dielectric function using this approach has the form:

$$\varepsilon_{eff}^{1/3} = f \varepsilon_m^{1/3} + (1-f) \varepsilon_h^{1/3}$$

The dielectric functions provided by the LLL model and the calculated normal incidence reflectance are shown in Supporting Figure 6. Since this model is dedicated to low volume fractions, percolating networks do not provide an adequate description of the measured spectra, which belong to high volume fraction percolating networks.

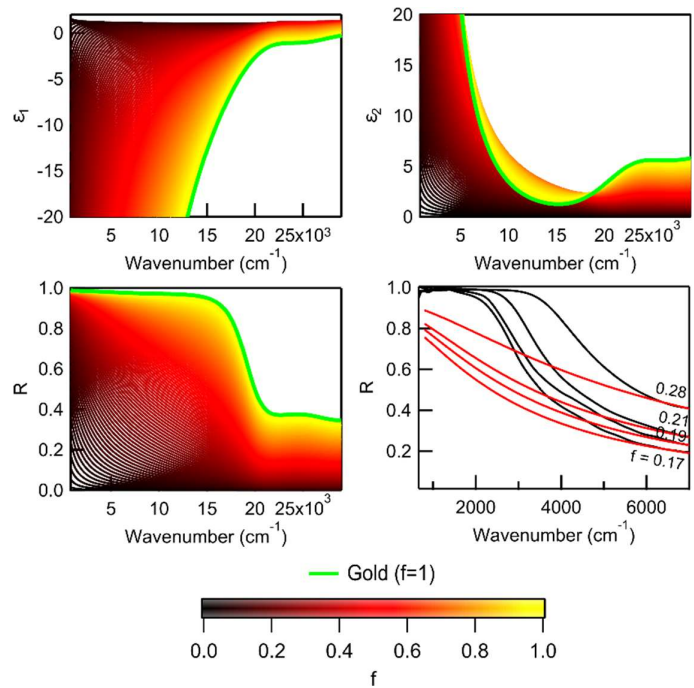

**Supporting Figure 6.** Fit of the optical response by means of the LLL EMA model.

**Supporting Refs**